\documentclass[aps,preprint,showpacs,showkeys,nofootinbib,floatfix]{revtex4}
\usepackage{epsfig}
\usepackage{amssymb}
\usepackage{graphicx}

\begin{document}

\title{Unraveling the pattern of the $XYZ$ mesons}
\author{J. Vijande}
\affiliation{Departamento de F\'{\i}sica At\'{o}mica, Molecular y Nuclear, Universidad de Valencia (UV)
and IFIC (UV-CSIC), Valencia, Spain.}
\author{A. Valcarce}
\affiliation{Departamento de F\'\i sica Fundamental, Universidad de Salamanca, E-37008
Salamanca, Spain}
\date{\today}

\begin{abstract}
We present a plausible mechanism for the origin of the $XYZ$ mesons in the heavy meson spectra 
within a standard quark-model picture.
We discuss the conditions required for the existence of four--quark bound states or resonances contributing to
the heavy meson spectra, being either compact or molecular. We concentrate on charmonium and bottomonium 
spectra, where several new states, difficult to understand as simple quark-antiquark pairs, have 
been reported by different experimental collaborations. The pivotal role played by entangled 
meson-meson thresholds is emphasized.
\end{abstract}

\maketitle

One of the most basic problems of QCD is to identify all the clusters of quarks,
antiquarks and gluons that are sufficiently bound by QCD interactions that they
are either stable particles or appropriately long-lived to be observed as resonances~\cite{Bra13}.
To this respect, charmonium spectroscopy has become a cornerstone during the ten years elapsed since 
the discovery of the first of the so-called $XYZ$ states, the $X(3872)$~\cite{Cho03}. 
Before this discovery, and based on Gell-Mann conjecture~\cite{Gel64}, the hadronic experimental 
data were classified either as $q\overline{q}$ or $qqq$ states according to $SU(3)$ irreducible 
representations. However, since 2003 more than twenty newly observed meson resonances reported 
by different experimental collaborations appeared, close to a two-meson threshold,
presenting properties that make a simple quark-antiquark structure unlikely~\cite{Bod13}. Although
this observation could be coincidental due to the large number of thresholds in the energy region
where the $XYZ$ mesons have been reported, it could also point to
a close relation between some particular thresholds and resonances contributing to the standard
quark-antiquark heavy meson spectroscopy. 
Several alternatives have been proposed in the literature to address these $XYZ$ states, being the most 
common ones conventional quarkonium~\cite{Eic04}, meson-meson molecules~\cite{Tor94}, quarkonium 
hybrids~\cite{Clo05}, four-quark states~\cite{Vij07a}, and dynamically generated 
resonances~\cite{Gam10}. However, none of the models that has been proposed provide a plausible
pattern for all the $XYZ$ mesons that have been observed.

In this letter we analyze heavy meson spectroscopy beyond open flavor thresholds considering higher order 
Fock space components looking for a general pattern of the $XYZ$ states. We will highlight the pivotal
role played by meson-meson thresholds in heavy meson spectroscopy 
and under which conditions, if any, they could cause a resonance to 
appear. This study is an interesting, and not settled, issue to address that may provide 
with a pattern to be compared with experiment and lattice QCD calculations. The outline is the following.
We will first describe the main characteristics of the quark model and the four-body formalism 
chosen to illustrate our findings. Then, we will analyze the thresholds open for 
four-quark states contributing to heavy meson spectroscopy and their main model-independent properties. 
Finally, we will show how these thresholds may entangle so a bound four-quark state or a resonance may emerge and 
the conditions required for that.

Although the conclusions of this study aim to be independent of the formalism selected to solve the four-body Schr\"odinger 
equation and the particular details of the quark model used, we have to choose a particular 
method, that is going to be the Hypherspherical Harmonic formalism (HH)~\cite{Bar06}, and model, the constituent quark 
cluster model (CQCM)~\cite{Vij05}, to illustrate them.
Within the HH method, the four--quark wave function is written as a sum of outer products of color, 
isospin, spin and radial terms to obtain basis functions that have well--defined 
symmetry under permutations of the quark pairs. By using hyperspherical coordinates one can write the Laplace 
operator, where the hyperspherical, or grand angular momentum, is the $9$--dimensional analogous of the 
angular momentum operator associated with the $3$--dimensional Laplacian~\cite{Fab83}. In the following we 
shall use the notations $K$ for the total hyperangular momentum. 
The CQCM model was proposed in the early 90's in an attempt to obtain a simultaneous description of the 
nucleon-nucleon interaction and the baryon spectra~\cite{Val05}. It was later on generalized to all 
flavor sectors giving a reasonable description of the meson and baryon spectra. 
The model is based on the assumption that the light--quark constituent mass appears because of the 
spontaneous breaking of the original $SU(3)_{L}\otimes SU(3)_{R}$ 
chiral symmetry at some momentum scale. In this domain of momenta, quarks interact through Goldstone 
boson exchange potentials. QCD perturbative effects are taken into account through the 
one-gluon-exchange potential. Finally, it incorporates confinement as dictated by unquenched 
lattice calculations. A detailed discussion of the model can be found in Refs.~\cite{Vij05,Val05}.

Standard mesons ($q\bar q$) and baryons ($qqq$) are the only clusters of quarks where it is not possible to construct a 
color singlet using a subset of their constituents. Thus, $q\bar q$ and $qqq$ states are proper solutions 
of the two- and three-quark hamiltonian, respectively, corresponding in all cases to bound states. 
This, however, is not the case for multiquark combinations, and in particular for four--quark states
addressing the meson spectra. Thus, when dealing with higher order Fock space contributions to
meson spectroscopy, one has to discriminate between possible four--quark bound states or resonances 
and simple pieces of the meson--meson continuum. For this purpose, one has to analyze the two--meson 
states that constitute the threshold for each set of quantum numbers.
These thresholds has to be determined assuming quantum number conservation within exactly the same 
scheme (parameters and interactions) used for the four--body calculation. If other models, parametrizations 
or experimental masses are used, then four-quark states might be misidentified as members
of the meson spectra while being simple pieces of the continuum. Working with strongly 
interacting particles, two--meson states should have well--defined total angular momentum ($J$) 
and parity ($P$). If the two mesons are identical, the spin--statistics theorem imposes a properly 
symmetrized wave function. Moreover, $C-$parity should be conserved in the final two--meson state for those 
four--quark states with well--defined $C-$parity. Finally, if noncentral forces are not considered, orbital angular 
momentum ($L$) and total spin ($S$) are also good quantum numbers. 

Given a general four--quark state, ($q_1q_2\bar q_3\bar q_4$), two different thresholds are 
allowed, $(q_1\bar q_3)(q_2\bar q_4)$ and $(q_1\bar q_4)(q_2\bar q_3)$. If the four--quark system 
contains identical quarks, like for instance $(QQ\bar n\bar n)$ (in the following $n$ 
stands for a light quark and $Q$ for a heavy $c$ or $b$ quark), the two thresholds are
identical, i.e., $(Q \bar n)(Q\bar n)$. The importance of this particular feature lies on the fact that 
a modification of the four--quark interaction would not necessarily translate into the 
mass of the two free-meson state. Therefore, the unique necessary condition required to have 
a four--quark bound state would be the existence of a sufficiently attractive interaction between 
quarks that do not coexist in the two free-meson states. This hypothesis was 
demonstrated by means of the Lippmann--Schwinger formalism in Ref.~\cite{Car11}, concluding the 
existence of a single stable isoscalar doubly charmed meson with quantum numbers $J^P=1^+$. 
    
For those cases containing a heavy quark and its corresponding heavy antiquark $(Q n \bar Q \bar n)$ the 
situation is remarkably different. Two different thresholds are allowed, namely $(Q\bar Q)(n\bar n)$ and $(Q\bar n)(n\bar Q)$. 
It has been proved~\cite{Bert80} that ground state solutions of the Schr\"odinger ($q_1\bar q_2$) two--body problem 
are concave in $m_{q_1}^{-1}+m_{q_2}^{-1}$ and hence $M_{Q\bar n}+M_{\bar Q n}\geqslant M_{Q\bar Q}+M_{n\bar n}$. 
This property is enforced both by nature\footnote{$M_{D^*}+M_{\bar D^*}=4014$ MeV $\geqslant M_{J/\psi}+M_{\omega}=3879$ MeV} 
and by all models in the literature unless forced to do otherwise. 
Although this relation among ground-state masses makes the assumption of a strictly flavor
independent potential, one should bear in mind that ground states of heavy mesons are perfectly
reproduced by a Cornell-like potential~\cite{Clo03}, that it is flavor independent.
The color-spin dependence of the potential would go in favor of this relation
for ground states (spin zero) because the color-spin interaction is attractive
for spin zero and comes suppresed as $1/(m_im_j)$, making even lighter the mesons
on the right hand side. Regarding the spin independent part,
the binding of a coulombic system is proportional to the reduced mass of the
interacting particles. Thus, for a two-meson threshold with a heavy-light light-heavy 
quark structure, the binding of any of the two mesons is proportional to the reduced 
mass of each meson, being close
to the mass of the light quark. However, if the two-meson state presents a
heavy-heavy light-light quark structure, the binding of the heavy-heavy meson increases 
proportionally to the mass of the heavy particle while that of the light-light meson remains constant,
becoming this threshold lighter than the heavy-light light-heavy two-meson structure.
Thus, it implies that in all relevant 
cases the lowest two-meson threshold for any $(Q n \bar Q \bar n)$ state will be the one made of quarkonium-light 
mesons, i.e., $(Q\bar Q)(n\bar n)$ (see Fig. 1 of Ref.~\cite{Car12}).
The interaction between the heavy, $(Q\bar Q)$, and light, $(n\bar n)$, mesons forming the lowest threshold is 
almost negligible, due to the absence of a light pseudoscalar exchange mechanism between them~\cite{Tor94}. 
Hence, any attractive effect in the four--quark system must have its origin in the interaction of the higher channel $(Q\bar n)(n\bar Q)$
or due to the coupled channel effect of the two thresholds, $(Q\bar Q)(n\bar n)\leftrightarrow(Q\bar n)(n\bar Q)$~\cite{Lut05}.
In Ref.~\cite{Vij07a} a comprehensive analysis of the $c\bar c n\bar n$ spectra was undertaken within the same working
framework we use in this letter. All 
isoscalar states with total orbital angular momentum $L\le 1$ were considered. No bound 
state was observed for any set of quantum numbers in any of the quark models considered. In all cases 
the four--quark system evolved for large values of $K$ to a well separated $(Q\bar Q)$--$(n\bar n)$ two--meson state, 
corresponding to the lowest threshold. We have repeated the same calculation enlarging the 
number of states to include both the isovector and bottom  sectors to no avail. No compact four-quark state 
was found for any isospin combination, heavy quark mass, or quark model studied.

In those energy regions where bound and unbound solutions of the four-quark hamiltonian coexist, methods based on 
infinite expansions become inefficient to hunt a bound state close to an unbound solution, because too many basis 
states would be required to disentangle them. The most 
interesting case where this may happen is in the vicinity of a two--meson threshold, because 
both the two free--meson state and a feasible slightly bound four--quark state are solutions of the same 
hamiltonian. Such cases have been studied by means of the Lippmann-Schwinger equation in Ref.~\cite{Car09},
looking at the Fredholm determinant $D_F(E)$ at zero 
energy~\cite{Gar87}. If there are no interactions then $D_F(0)=1$, 
if the system is attractive then $D_F(0)<1$, and if a
bound state exists then $D_F(0)<0$. All states made of $S$ wave 
($Q\bar n$)--($n\bar Q$) mesons up to $J = 2$ were scrutinized. A few channels 
were found to be slightly attractive, 
$D\bar{D}$ with $(I)J^{PC}=(0)0^{++}$, $D\bar{D}^*$ with $(0)1^{++}$ 
and $D^*\bar{D}^*$ with $(0)0^{++}$, $(0)2^{++}$, and $(1)2^{++}$, close 
to the results of Ref.~\cite{Tor94}. However, the only bound state appeared in the 
$(I)J^{PC}=(0)1^{++}$ channel as a consequence of the coupling between 
$D\bar D^*$ and $J/\Psi \omega$ two--meson channels. 

The conclusions of Refs.~\cite{Vij07a} and~\cite{Car09} point to a convoluted four--quark 
molecular structure with a dominant $D\bar D^*$ component for the $X(3872)$. However, this is not the only supernumerary state that 
has appeared in the charm and bottom sectors during the last years. A comprehensive list of 
such $XYZ$ mesons and their properties can be found in Ref.~\cite{Bod13}. 
20 states have been reported by different experimental collaborations in the charmonium sector 
above the $D\bar D$ threshold, 15 of them neutral and 5 charged. In the bottom sector 2 charged 
and 1 neutral state have been reported. Of those 23 states only 8 have been observed independently 
by two different collaborations and with significance greater than 5$\sigma$: $X(3872)$, $X(3915)$, $\chi_{c2}(2P)$, $G(3900)$, $Y(4140)$, $Y(4260)$, 
$Y(4360)$, and $Z_c^+(3900)$ (see Table I of Ref.~\cite{Bod13}). Therefore, although some of them might not resist cross-check 
examination by independent experimental collaborations, others are clearly established as 
real resonances and therefore they have to be accounted for in any description of the meson spectra.
While some of them, like the $\chi_{c2}(2P)$, seem to fit nicely within a naive quark-antiquark scheme, others do not. 

\begin{figure}[tb]
\begin{center}
\epsfig{file=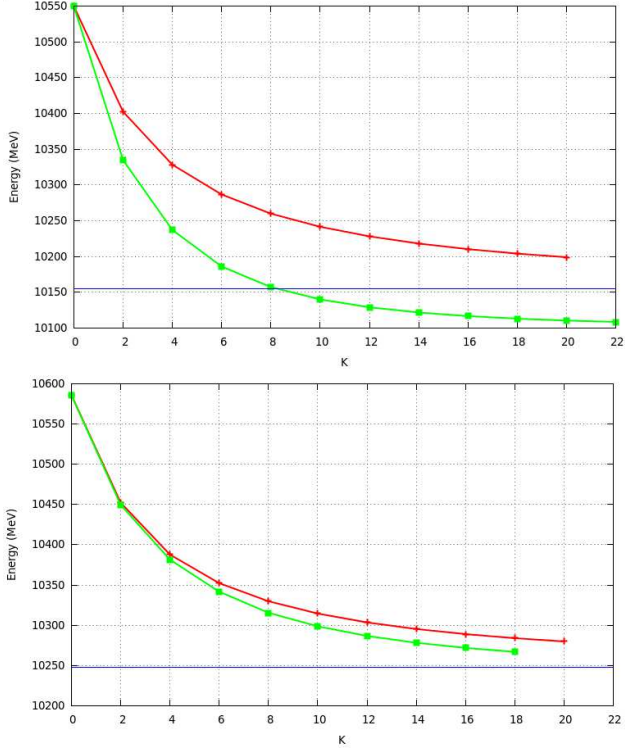, height=10cm}
\end{center}
\caption{Convergence of $b\bar bn\bar n$ with quantum numbers $L=0$, $S=1$, $C=+1$ $P=+1$, $I=0$ (upper panel) 
and $I=1$ (lower panel). Red lines correspond to the case where the thresholds are non-degenerate and green 
lines to the case where they are almost degenerate.}
\label{fig1}
\end{figure}

When four-quark components are considered in the wave function of charmonium, there are 72 $c\bar c n\bar n$ 
combinations of quantum numbers for total orbital angular momentum $L<3$. Therefore, the question is not 
whether it is possible to design a model, or a formalism, able to match one of the newly observed $XYZ$ states 
with a particular set of these quantum numbers but to understand where the attraction comes from 
and to explain the systematic that predicts where, if anywhere, 
experimentalists and theoreticians alike should look into.
Since the lowest threshold interaction is rather weak, the possibility to obtain
bound states may only stem from the vicinity of an attractive $(c\bar n)(n\bar c)$ threshold 
coupled sufficiently as to bind the system as it occurs with the X(3872)~\cite{Car09}. 
Although for heavier mesons interactions are more attractive~\cite{Car12}, the effect of
channel coupling may not be enough to favor binding. Thus,
to check the efficiency of this mechanism, let us solve with the HH formalism 
the bottom counterpart of the $X(3872)$, $b\bar bn\bar n$ with quantum numbers 
$L=0$, $S=1$, $I=0$, $C=+1$, and $P=+1$. Within the CQCM the corresponding lowest 
thresholds, $B\bar B^*$ (10611 MeV) and $\Upsilon\omega$ (10155 MeV), are 456 MeV apart. We 
show in Fig.~\ref{fig1}(upper panel) the convergence pattern of the energy of the
four-quark system as a function of the 
hyperangular momenta $K$. It can be clearly seen how the energy of the four--quark system (red line)
is converging to the lowest threshold $\Upsilon\omega$ (horizontal blue line), what is a sharp signal
of an unbound state. One could however play around with the model parameters 
to almost degenerate both thresholds by adding 
attraction in the heavy-light $bn$ sector\footnote{We have slightly increased
the $\alpha_s(bn)$ strong coupling constant from 0.55 to 0.85, what would move the gap
between thresholds without changing the lowest threshold.}, what would also increase the coupled channel effect
strengthening the $B\bar B^*\leftrightarrow \Upsilon\omega$ transition interaction. 
When this is done we note (green line) 
that the energy drops below threshold and a bound state emerges\footnote{Additional 
tests on the wave function, not included here for the sake of brevity, following 
the formalism outlined in Ref.~\cite{Vij09c} have been performed to verify that this state is a real bound state.}. One may wonder if only the close-to-degeneracy of the
thresholds is sufficient to bind this type of four--quark systems. If this would be the case, then the charged partner 
of this four--quark state ($I=1$) should behave exactly in the same manner. However, amazingly this is not so. 
In Fig.~\ref{fig1}(lower panel) we depict the convergence of the isovector state 
as a function of $K$ for both cases, non-degenerate thresholds (red line) 
and almost degenerate ones (green line). In this case the lowest threshold would be $\Upsilon\rho$ (10248 MeV). It can be observed that in both cases the 
four--quark state converges to the lowest threshold and does not form a bound state.

\begin{figure}
\begin{center}
\epsfig{file=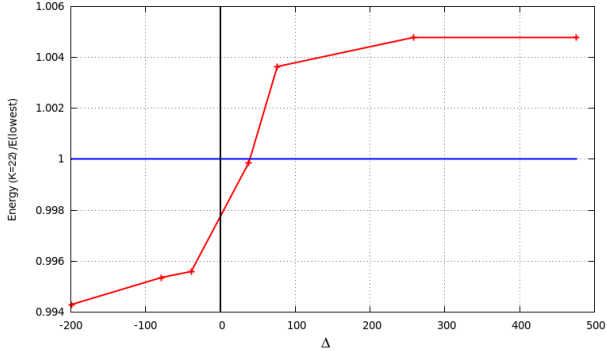, width=8cm}
\end{center}
\vspace*{-0.5cm}
\caption{Energy divided by the lowest two-meson threshold of $b\bar bn\bar n$ with quantum 
numbers $L=0$, $S=1$, $C=+1$ $P=+1$, $I=0$ as a function of $\Delta = E[(b\bar n)(n\bar b)]-E[(b\bar b)(n\bar n)]$.}
\label{fig2}
\end{figure}

Thus, when the $(Q\bar Q)(n\bar n$) and $(Q\bar n)(n\bar Q)$ thresholds are sufficiently far away, which means that the interaction in the higher
$(Q\bar n)(n\bar Q)$ state is weak and therefore the coupled channel effect small, no bound states are found for any set of parameters. 
However, when the thresholds move closer, i.e., the attraction in the higher two-meson state
and the coupled channel strength are simultaneously increased, bound states may appear for a subset of 
quantum numbers. Hence, threshold vicinity is a required but not sufficient 
condition to bind a four--quark state. An additional condition is 
required to allow the emergence of such bound states. Such condition is the 
existence of an attractive interaction in the higher $(Q\bar n)(n\bar Q)$ 
two--meson system that would also give rise to a strong $(Q\bar Q)(n\bar n)\leftrightarrow(Q\bar n)(n\bar Q)$ coupling. 
To neatly illustrate this conclusion we return to the four--quark 
state $b\bar bn\bar n$ with quantum numbers $L=0$, $S=1$, $I=0$, 
$C=+1$, and $P=+1$. In Refs.~\cite{Car09,Car12} it was proved that the interaction provided 
by the CQCM model is attractive for these quantum numbers in the 
charm sector, although does not present a $D\bar D^*$ bound state. In the bottom sector the attraction is enhanced~\cite{Car12}. Thus, we have solved 
the four--body problem as a function of the threshold energy difference, 
$\Delta = E[(b\bar n)(n\bar b)]-E[(b\bar b)(n\bar n)]$, ranging from 500 MeV to $-$200 MeV. 
We show in Fig.~\ref{fig2} the energy normalized to the mass of the lowest two-meson threshold for each particular case, 
i.e., values smaller than 1 will point to a bound state and those larger to an unbound state. 
In this case, the results can be separated into two distinct categories: (i) $\Delta\gtrsim 50$ MeV 
and (ii) $\Delta\lesssim 50$ MeV. When the thresholds are separated by more than 50 MeV, 
the attractive interaction in the $B\bar B^*$ system and the coupled channel effect is not sufficient to overcome 
the threshold energy gap, and therefore the four--quark system evolves to an unbound two--meson state. 
However, when the thresholds get closer, even reversed for $\Delta<0$, the system 
becomes a compact four--quark state. Of particular interest are those cases 
where $\Delta\simeq50$ MeV in Fig.~\ref{fig2}. In this case the attraction in the higher channel together
with the coupled channel effect barely overcomes 
the threshold energy difference, and hence its wave function becomes strongly
entangled. This would generate a molecular state just close to threshold.
It should also be noted that $\Delta_0$, the $\Delta$ value for which Energy($4q$)/E(lowest) is equal to one, will change depending on 
the particular set of quantum numbers considered. However, $\Delta_0$ = 0 implies that no binding energy is provided by the upper threshold 
and the off-diagonal terms and therefore such solution will not correspond to a bound state. 
In that case  our calculation will be simply providing a piece of the meson-meson unbound continuum.

As one can see the mechanism proposed is restrictive enough as not
predicting a proliferation of bound states when explaining the existence
of an hypothetical molecular structure. This mechanism may also work in other two-hadron
systems. For the sake of completeness, let us suggest a possible scenario
where it could be relevant for an exotic meson-baryon system in the charm baryon spectroscopy.
There is an unexplained structure recently reported by the BABAR Collaboration~\cite{Lee12}, 
with a mass of 3250 MeV/c$^2$ in the $\Sigma_c^{++} \pi^- \pi^-$ invariant mass
that may be a consequence of the close-to-degeneracy of the lowest thresholds with
$I=2$ and $J^P=5/2^-$,  $\Delta D^*$ and $\Sigma^*_c \rho$ and the attractive interaction
of the $\Delta D^*$ system~\cite{Car14}. 
Such state was identified by QCD sum rules analyses~\cite{Alb13} as a pentaquark candidate, the $\Theta_c(3250)$.

To summarize, we have presented a plausible mechanism for the origin of the $XYZ$ mesons in the 
heavy meson spectra within a standard quark-model picture. The existence of open flavor 
two--meson thresholds is a feature of the heavy meson spectra that needs to be considered 
as a relevant ingredient into any description of the plethora of new states reported in 
charmonium and bottomonium spectroscopy. They might be, at a first glance, identified 
with simple quark-antiquark states, however in some cases their energies and decay 
properties do not match such oversimplified picture. 
Our results prove the relevance of higher order Fock space components
through the allowed two-meson thresholds. On the one hand, 
one has the lower $(Q\bar Q)(n\bar n)$ system, made 
by almost noninteracting mesons, that constitutes the natural breaking apart end-state. On the other hand, the 
higher $(Q\bar n)(n\bar Q)$ system appears. {\it When there is an attractive interaction characterizing the 
$(Q\bar n)(n\bar Q)$ upper system combined with a strong enough $(Q\bar Q)(n\bar n)\leftrightarrow(Q\bar n)(n\bar Q)$
coupling, together with the vicinity of the two allowed thresholds, 
a four--quark bound state may emerge.}
This is a rather restricting property that makes that no $c\bar cn\bar n$ 
or $b\bar bn\bar n$ bound state was reported in the analysis performed in 
Ref.~\cite{Vij07a}. Thus, an analysis of the threshold energy difference 
and the strength of the $(Q\bar n)(n\bar Q)$ and  $(Q\bar Q)(n\bar n)\leftrightarrow(Q\bar n)(n\bar Q)$
interactions using different state-of-the-art 
quark models for all allowed sets of quantum numbers is required to predict whose 
sets are candidates to lodge one of the $XYZ$ states. 

Once this is performed, the present
experimental effort with ongoing experiments at BESIII, current analyses by the LHC collaboration
and future experiments at Belle II and Panda together with the very impressive results that are 
being obtained by lattice gauge theory calculations~\cite{Liu12} may confirm the theoretical 
expectations of our quark-model calculation pattern that will provide with a deep understanding 
of low-energy realizations of QCD.

\acknowledgments
This work has been partially funded by the Spanish Ministerio de Educaci\'on y Ciencia and EU FEDER under Contracts No. FPA2010-21750 and FPA2013-47443,
by the Spanish Consolider-Ingenio 2010 Program CPAN (CSD2007-00042) and by Generalitat Valenciana Prometeo/2009/129. A.V. thanks finantial support from the Programa
Propio I of the University of Salamanca.

\end{document}